\newcommand{\be}{\begin{equation}}
\newcommand{\ee}{\end{equation}}

\documentstyle[epsf]{article}
\def\boxit#1{\vbox{\hrule\hrule\hrule\hbox{\vrule\vrule\vrule\kern3pt\vbox{\kern3pt#1\kern3pt}\kern3pt\vrule\vrule\vrule}\hrule\hrule\hrule}}
\def\comma{ \hspace{2mm}, }
\def\period{ \hspace{2mm}. }

\def\b{ \hspace{2mm} }

\def\di{ \partial }
\def\de{ {\rm d} }
\def\half{ \frac{1}{2} }
\def\minus{ \mbox - \, }
\def\const{ 8\pi G }
\def\div{ \partial_v }
\def\dir{ \partial_r }

\begin{document}
\onecolumn
\begin{flushright}
WATPHYS TH-94/06\\
gr-qc/yymmddd
\end{flushright}
\vfill
\begin{center}
{\Large \bf
Interior Structure of a Charged Spinning Black Hole in $(2+1)$-
Dimensions}\\
\vfill
J.S.F. Chan$^{(2)}$, K.C.K. Chan$^{(1)}$ and R.B. Mann$^{(1,2)}$\\
\vspace{2cm}
(1) Department of Physics, University of Waterloo, Waterloo, Ontario, Canada
N2L 3G1\\
(2) Department of Applied Mathematics,
University of Waterloo, Waterloo, Ontario, Canada, N2L 3G1\\
\vspace{2cm}

\vfill
%
%

\begin{abstract}
The phenomenon of mass inflation is shown to occur for a rotating black
hole. We demonstrate this feature in $(2+1)$ dimensions by extending the
charged spinning BTZ black hole to Vaidya form. We find that the mass
function diverges in a manner quantitatively similar to its static
counterparts in $(3+1)$, $(2+1)$ and $(1+1)$ dimensions.
\end{abstract}

\vfill
\end{center}
\clearpage

\twocolumn

The endpoint of gravitationally collapsing matter is a subject of
long-standing interest in general relativity.  Given the plausible (but
unproven) hypothesis of cosmic censorship, it has been established that the
spacetime exterior to a collapsing body relaxes to that of a Kerr-Newman
(KN) black hole, with radiative perturbations decaying as advanced time
increases according to a power law.

The question of what happens to the collapsing matter, along with
everything else falling into the black hole, is somewhat more problematic.
Infalling matter will either encounter a spacelike region of diverging
curvature (at which point quantum gravitational effects presumably
dominate) or alternatively will avoid the singularity and emerge into
another universe via a `white hole', the KN geometry being prototypical of
this latter possibility.  However it has been shown that the interior
geometry of a KN black hole is unstable: the stress-energy associated with
massless test fields diverges at a null hypersurface inside the black hole
called the Cauchy horizon \cite{Penrose}.  Any object falling into a KN
black hole must eventually cross the Cauchy horizon, and so an
understanding of its stability is intimately connected to the question of
the final fate of the infalling matter.

Significant progress on this problem was made by  Poisson and Israel
\cite{Poisson}, who demonstrated that the Cauchy horizon of the
Reissner-Nordstr\"{o}m solution forbids any evolution of spacetime beyond
this horizon.  Inside the black hole the mass parameter becomes unbounded
due to the presence of ingoing and backscattered outgoing radiation,
and the Kretschmann scalar diverges.
This phenomenon was subsequently  confirmed by Ori \cite{Ori}, who
constructed an exact solution of the Einstein-Maxwell equations in a
simpler model. He argued that the mass inflation singularity was
too weak to forbid passage through the Cauchy horizon, since
its tidal forces do not necessarily destroy any physical
objects. This extensibility problem remains a subject of some controversy
\cite{Herman,Bonnano}. Mass inflation has also been shown to take place in
lower-dimensional analogs of the Reissner-Nordstr\"{o}m solution, both
in $(1+1)$ \cite{JChan,Droz} and $(2+1)$ dimensions \cite{Husain}.

As the causal structure of the Reissner-Nordstr\"{o}m  geometry is
similar to that of the KN geometry, it is generally
believed that the KN Cauchy horizon forms a similar obstruction to the
evolution of the spacetime. However no models have appeared to date.

We present here the first exact mass-inflation solution of a charged
spinning black hole. In particular, we consider the $(2+1)$-dimensional BTZ
black hole geometry \cite{BTZ} in the context of Ori's model,  and show
explicitly that it mass-inflates in a manner analogous to its
Reissner-Nordstr\"{o}m  counterpart.

Let us consider Einstein's equations with cosmological constant
  $\Lambda < 0$ in $(2+1)$ dimensions:
  \begin{eqnarray}
    G_{\mu \nu} + \Lambda\,g_{\mu \nu} & = &
    \const\,T_{\mu \nu} \comma \label{E1}
  \end{eqnarray}
  where $\const$ is the coupling constant in $(2+1)$ dimensions and
the stress-energy tensor is that for the electrovacuum.  The
  standard BTZ spinning black hole
  solution for (\ref{E1}) with static electric charge has the form
  \begin{equation}
    \de s^2 =
    \minus N^2(r)\de t^2 + N^{\minus 2}\de r^2
    + r^2\left(N^\phi\de t + \de \phi\right)^2 \label{E2}
  \end{equation}
where $N^\phi := \minus \frac{J}{2\,r^2}$ and
  \begin{eqnarray}
    N^2(r)  :=  \minus \Lambda\,r^2 - M + \frac{J^2}{4\,r^2}
    - 4\pi G q^2\,\ln(r/r_0)  \period \label{E3}
  \end{eqnarray}
  The integration constants $M$ and $J$ are interpreted as the mass
  and the angular momentum of the black hole respectively
  \cite{Cangemi,Jolien}
  and the constant $q$ is the charge carried by the hole.
  Figure 1 shows the causal structure of the BTZ spacetime.
\begin{center}
\leavevmode
\hbox{\epsfxsize=7.5cm \epsfysize=11.0cm \boxit{\epsffile{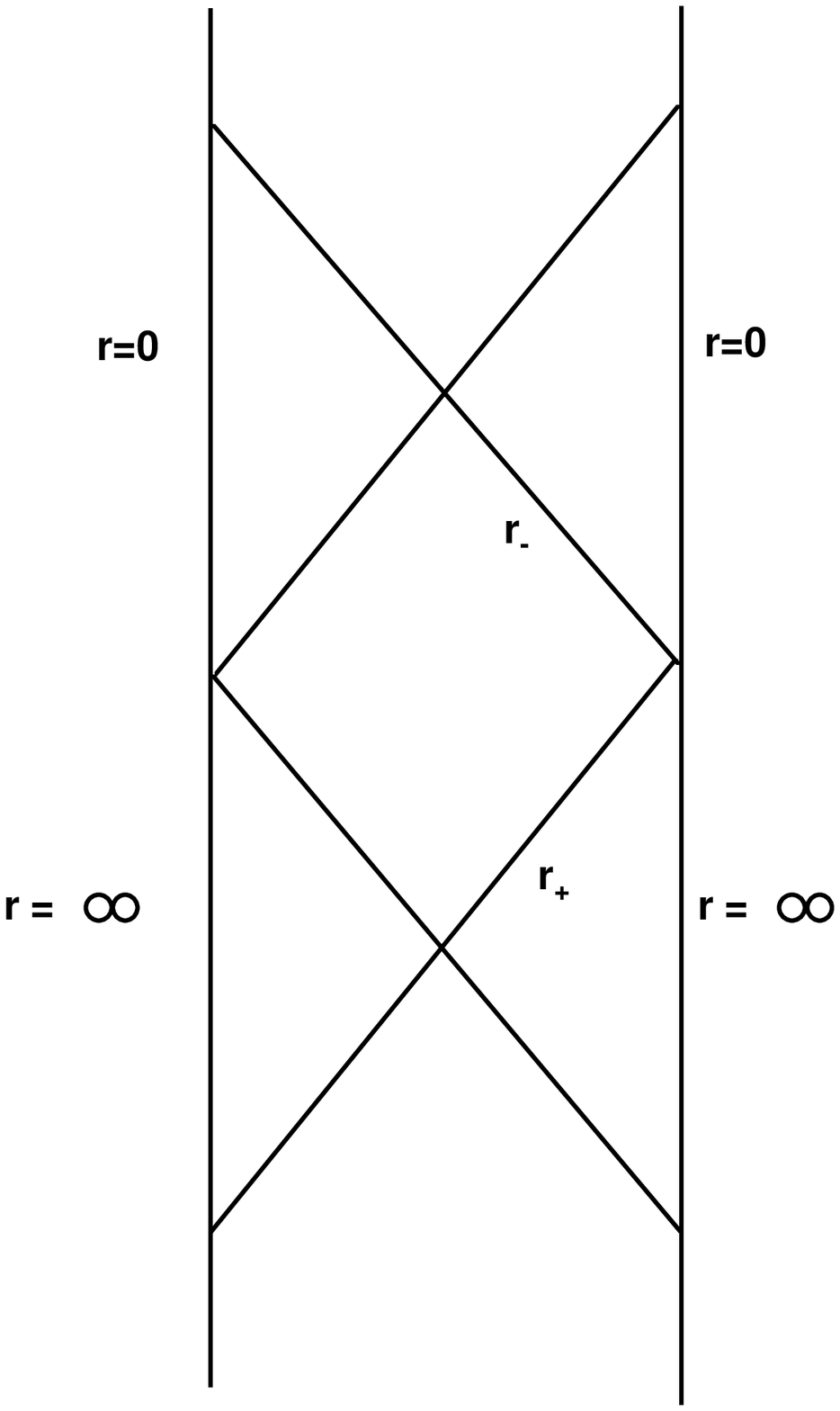}}}
\parbox{6.5cm}{\small Fig. 1. Penrose diagram for the BTZ solution}
\end{center}
There are two horizons, an outer horizon at $r=r_+$ and an inner
(Cauchy) horizon at $r=r_-$, where
$$
  r^2_{\pm} \equiv
  \frac{M}{2|\Lambda|}\left(1 \pm \sqrt{1-|\Lambda|\frac{J^2}{M^2}}
  \right)
$$
when $q=0$. In this case there is
a singularity in the causal structure \cite{BTZ};
however when this solution is matched to a collapsing cloud of dust,
there is again a curvature singularity at $r=0$ \cite{Ross}.
For non-zero $q$, $r_{\pm}$ are the solutions to the
transcendental equation $N(r_{\pm})=0$, and there is a curvature
singularity at $r=0$.

Defining new coordinates $v$ and $\theta$ as
  \begin{equation}
    v  :=  t + \int^r \frac{\de \tilde{r}}{N^2(\tilde{r})}
    \quad {\rm and} \quad
\theta :=  \phi - \int^r \frac{N^\phi(\tilde{r})}{N^2(\tilde{r})}\de \tilde{r}
    \label{E4}
  \end{equation}
the metric can be written as
\begin{eqnarray}
    \de s^2 &=& \minus \left[\,{\cal M}(r) -
    M - \frac{J^2}{4\,r^2}\right]\,\de v^2 \nonumber\\
    &&\quad + 2\de v\de r
    - J\de v\de \theta + r^2\de \theta^2  \label{E5}
  \end{eqnarray}
where the function ${\cal M}(r) =
    \minus \Lambda\,r^2 + \frac{J^2}{4\,r^2}
    - 4\pi G q^2\ln(r/r_0)$.
  In this coordinate system, $\div$ and $\di_{\theta}$ are Killing
  vectors. As one approaches the
  Cauchy horizon, $r$ is decreasing and $N^2$ approaches zero
  from below; thus $v$ tends to positive infinity.

Consider next the stress-energy-momentum tensor of null spinning
  dust which has the form
  \begin{eqnarray}
    \left[\,{\cal T}_{\mu \nu}\,\right] & = &
    \left[\,\begin{array}{ccc}
    \hat{\rho}(v,r)          & 0 & \minus \hat{\omega}(v,r) \\
    0                        & 0 & 0 \\
    \minus \hat{\omega}(v,r) & 0 & 0
    \end{array}\,\right]  \label{E7}
  \end{eqnarray}
where the energy density $\hat{\rho}$ and angular momentum density
$\hat{\omega}$ have the form
  \begin{equation}
    \hat{\rho}(v,r) =
    \frac{\rho(v)}{r} + \frac{j(v)\,\omega(v)}{2\,r^3} \quad{\rm and}
    \quad
    \hat{\omega}(v,r) =
    \frac{\omega(v)}{r} \label{E8}
  \end{equation}
as may easily be shown from the conservation laws.
Incorporating the
stress-energy-momentum tensor for the Maxwell field along with (\ref{E7})
into Einstein's equations (\ref{E1}) we obtain an exact solution
\begin{equation}
    \de s^2
    = \minus \alpha(r,v)\de v^2 + 2\de v\de r
    - j(v)\de v\de \theta + r^2\de \theta^2  \label{E9A}
  \end{equation}
with $\alpha := {\cal M}(r,v) -  m(v) - \frac{j^2(v)}{4\,r^2}$,
where now ${\cal M}(r,v) = \minus \Lambda\,r^2 + \frac{j^2(v)}{4\,r^2}
- 4\pi G q^2\ln(r/r_0) \equiv N^2(r,v) + m(v)$. The functions
$m(v)$ and $j(v)$ satisfy the ordinary differential equations
  \begin{eqnarray}
    \frac{dm(v)}{dv} = 16\pi G \rho(v) \quad {\rm and} \quad
    \frac{dj(v)}{dv}= 16\pi G \omega(v) \period \label{E10}
  \end{eqnarray}
When $j = 0$ this solution is analogous to the charged Vaidya solution
and represents a black hole which is irradiated by an influx of null
radiation \cite{Husain}.
When $j$ is non-zero, the black hole and the surrounding null
dust rotate. Uniform rotation takes place only when $j$ is a non-zero
constant.

We consider next a pulse of radiation along an outgoing null ring S. For
simplicity we take $\omega(v)=0$, so that $j(v) = J =$ constant.
As S is a ring
in $r$-$\theta$ coordinates, we denote the region enclosed by the ring as I
and its complement as region II, characterized by mass functions
$m = m_{1}(v_{1})$, and $m = m_{2}(v_{2})$ respectively.
We proceed now to match two patches of solution (\ref{E9A}), one from
each region, along the ring S.

Any null ray satisfies the equation
  \begin{equation}
    \minus \alpha\dot{v}^2(\lambda) + 2\dot{v}(\lambda)\dot{r}(\lambda)
    - J\,\dot{v}(\lambda)\dot{\theta}(\lambda)
    + r^2\,\dot{\theta}^2(\lambda) = 0  \label{E11}
  \end{equation}
  where $\lambda$ is an affine parameter and the dot denotes
  derivative with respect to $\lambda$. Without loss of generality,
  we choose the parameter $\lambda$ to be zero at the Cauchy horizon
  and positive beyond that. The geodesic equations are
  \begin{eqnarray}
    2\,\ddot{v} & = &
    \minus \dir \alpha\,\dot{v}^2 + 2\,r\,\dot{\theta}^2 \comma
    \label{E13} \\
    0 & = &
    \frac{\de}{\de \lambda} \left[\,\minus J\,\dot{v} +
2\,r^2\,\dot{\theta}\,\right] \label{E14}
  \end{eqnarray}
Equation (\ref{E14}) can be integrated and gives
$\dot{\theta} =  \frac{J}{2\,r^2}\,\dot{v}$, where we have set a constant
of the motion $g(\partial_\theta,u)=0$, $u^\mu$ being the 3-velocity of
the ring. Equation (\ref{E13}) then becomes
  \begin{equation}
    \frac{\de}{\de \lambda} \left[\,\frac{2}{\dot{v}}\,\right] =
    \partial_r{\cal M}(r) \label{E16}
\end{equation}
and the null condition (\ref{E11}) implies
  \begin{equation}
    2\,\dot{v}\,\dot{r} = N^2(r,v)\,\dot{v}^2 \period \label{E17}
  \end{equation}

Defining a function $R$ such that $2\,\pi\,R$ is the perimeter of
  the ring S and $z(\lambda) := 2\,R(\lambda) / \dot{v}(\lambda)$
it is straightforward to show that (\ref{E16}) and (\ref{E17}) yield
the matching conditions
\begin{equation}
    m_{i}(v_{i}(\lambda)) =
    {\cal M}(R(\lambda)) + R(\lambda){\cal M}'(R(\lambda))
    - \dot{z}_{i}(\lambda)  \label{E19A}
\end{equation}
\begin{eqnarray}
    v_{i}(\lambda) &=&
    2\,\int^\lambda R(\zeta) / z_{i}(\zeta)\,\de \zeta \comma
    \label{E19B} \\
    z_{i}(\lambda) &=&
    R(\lambda)\,\left[\,Z_{i} + \int^\lambda_0
    {\cal M}'(R(\zeta))\,\de \zeta\,\right] \period \label{E19C}
  \end{eqnarray}
where the subscript $i$ has a value either `1' or `2' to denote quantities
defined in the respective regions.  The terms $Z_{i}$ in equation
(\ref{E19C}) are integration constants. Given the boundary function $R$,
equations (\ref{E19A}) to (\ref{E19C}) determine the evolution of the
spacetime.

We define the ``mass'' of the ring as
  \begin{equation}
    \Delta m(\lambda)  :=  m_{2}(\lambda) - m_{1}(\lambda) =
    (Z_{1} - Z_{2})\,\dot{R}(\lambda) \period \label{E20}
  \end{equation}
Moreover, we define a constant $M := m_{1}(\lambda)  + \delta m(\lambda)$
as the final mass of the black hole observed in region I after it has
absorbed all the ingoing radiation. The term $\delta m$ is interpreted as
the mass of the flux of ingoing radiation. Because the Cauchy horizon
corresponds to the limit $v_{1} \rightarrow \infty$, we expect that
  \begin{eqnarray*}
    \lim_{\lambda \rightarrow 0^{\minus}} \dot{v}_{1}(\lambda) \b = \b
    \frac{2}{Z_{1}} \b = \b \infty \b
  \end{eqnarray*}
implying $Z_{1} = 0$. Since $\dot{R}(\lambda)$ is expected to be negative
inside the black hole the sign of $Z_{2}$ must be positive in order to have
a positive-energy ring S.

Equation (\ref{E19C}) can be written as
  \begin{eqnarray}
    z_{i}(\lambda) & = &
    R(\lambda)\,\left[\,Z_{i} - 2\,k_o\,\lambda\,\right] \comma \label{E22}
  \end{eqnarray}
where $k_o$ is defined as
\begin{eqnarray}
    k_o & := & \minus \frac{1}{2}\,{\cal M}'(R(\epsilon)) \label{E23}
\end{eqnarray}
  By the Mean
  Value Theorem, the small constant $\epsilon\,\in (\,\lambda,0\,)$.
  Since $N^2(r,v) = {\cal M}(r) - m(v)$, the slope of ${\cal M}$
  must be negative at Cauchy horizon (see Figure 2), and so
  $k_o$ is positive definite.
\epsfxsize=7.5cm
\epsffile{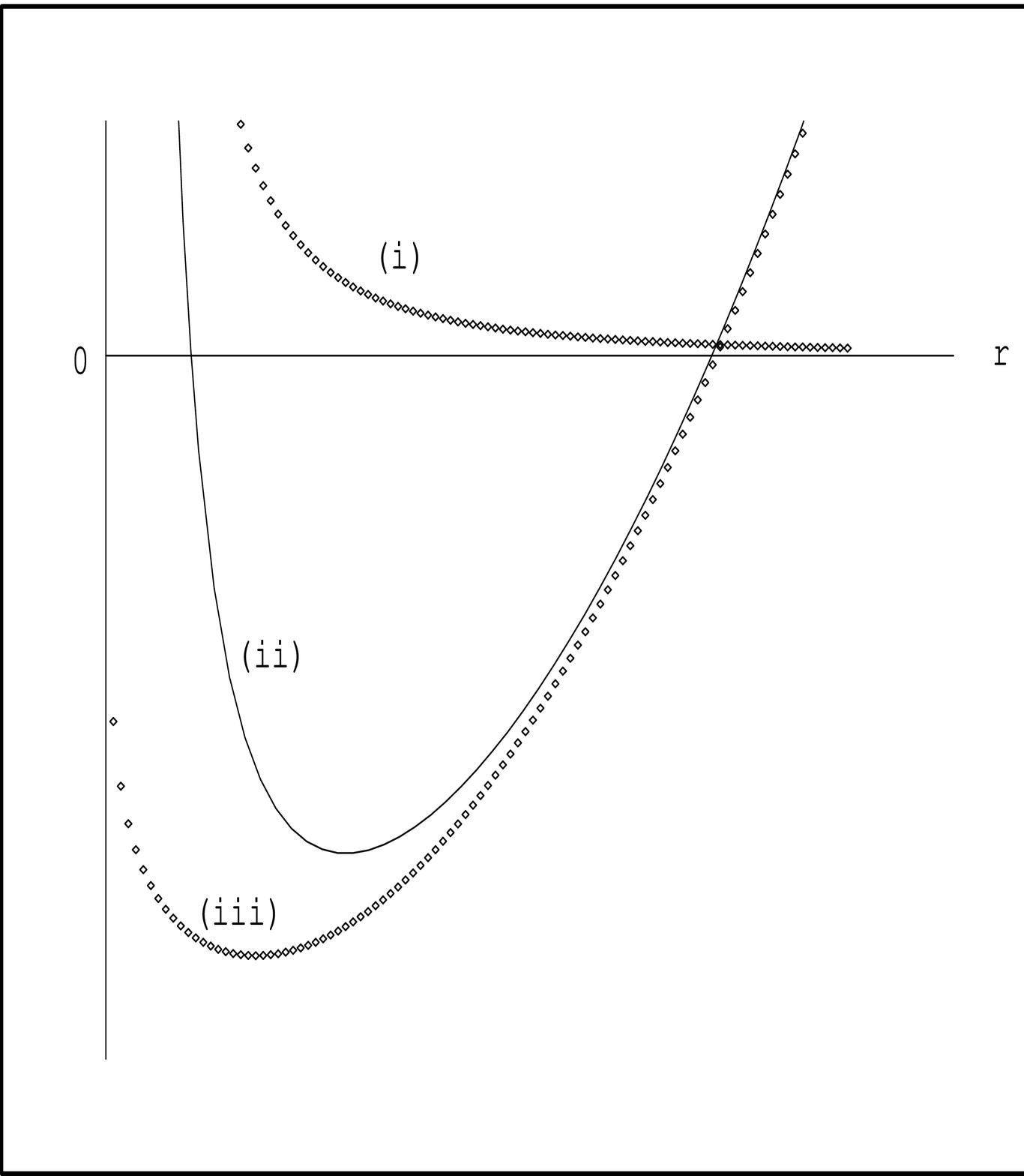}
\begin{center}
\parbox{6.5cm}{
\small
Fig. 2. The metric components: (i) $[r\,N^\phi]^2$, (ii) $N^2$,
(iii) $\alpha$}
\end{center}
\mbox{\hspace{1cm}}
\vspace*{0.1cm}\\

Equation (\ref{E17}) yields an approximation
  \begin{equation}
  \dot{R}(\lambda)
 \approx  \minus \half\left|\,M - m_{1}(v_{1}(\lambda))\right|
 \dot{v}_{1}(\lambda)
 =  \frac{1}{2\,k_o\,\lambda}\,\delta m(\lambda) \label{E24}
  \end{equation}
when $|\,\lambda\,| \ll 1$. On the other hand, since
\begin{eqnarray*}
    v_{i}(\lambda) \b = \b
    2\,\int^\lambda \frac{R(\zeta)}{z_{i}(\zeta)}\,\de \zeta \b = \b
    \int^\lambda \frac{2}{Z_{i} - 2\,k_o\,\zeta}\,\de \zeta \comma
  \end{eqnarray*}
  when the magnitude of $\lambda$ is small, $v_{i}$ can be approximated as
  \begin{eqnarray}
    v_{1}(\lambda) =
    \minus \frac{1}{k_o}\,\ln|\,\lambda\,| \quad{\rm and}\quad
    v_{2}(\lambda) \approx
    \frac{2}{Z_{2}}\,\lambda \period \label{E25}
  \end{eqnarray}

As a result, we obtain
  \begin{equation}
 m_{2}(v_2)  \approx
    M - h\left[\,1 + \frac{1}{k_o v_{2}}\right]
k_o^p\left|{\ln\left|\frac{Z_2 v_2}{2}\right|}\right|^{\minus p}
    \label{E26}
  \end{equation}
where a power law fall off $\delta m(\lambda) \sim h v_1^{-p}$
has been assumed \cite{Price}.
Thus $m_{2}$ diverges to positive infinity as $v_2\to 0^{-}$,
since  $Z_{2}$ and $k_o$ are positive.

We have shown that, at least in $(2+1)$ dimensions, all the generic
features of spherically symmetric mass inflation are preserved when
the black hole has angular momentum,
as is expected to be the case for $(3+1)$ dimensional black holes
\cite{Bonnano}.
We have checked the tidal forces associated with the mass inflation
singularity and have found that they produce a bounded distortion
of a $(2+1)$ dimensional physical object.  Furthermore, it
is straightforward to check that the Kretschmann
scalar is finite at the singularity, in contrast to the $(3+1)$-dimensional
case.  This suggests that classical continuation of the spacetime beyond
the mass-inflation singularity is not forbidden as Ori has
suggested\cite{Ori}; indeed, the $(2+1)$ dimensional model presented here
may permit an explicit construction of such a continuation.
\bigskip

This work was supported by the Natural Sciences
  and Engineering Research Council of Canada.


\begin{thebibliography}{9}

 \bibitem{Penrose} R. Penrose in {\em Battlle Rencontres}, edited by
		    C. M. De Witt and J. A. Wheeler,
		    New York: Benjamin (1968);
  \bibitem{Poisson} E. Poisson and W. Israel,
                    Phys. Rev. D {\bf 41}, 6, 1796 (1990).
  \bibitem{Ori} A. Ori, Phys. Rev. Lett. {\bf 67}, 7, 789 (1991).
  \bibitem{Herman} R. Herman and W. A. Hiscock,  Phys. Rev. Lett. {\bf 46},
		   4, 1863-1865 (1992).
  \bibitem{Bonnano} A. Bonanno, S. Droz, W. Israel and S. Morsink,
  preprint gr-qc/9403019  Alberta-Thy-9-94.
  \bibitem{JChan} J.S.F. Chan and R.B. Mann,
  WATPHYS-TH94/04, gr-qc/9406021.
  \bibitem{Droz} S. Droz, University of Alberta preprint (1994).
  \bibitem{Husain} V. Husain, preprint gr-qc/9404047  Alberta-Thy-12-94.
  \bibitem{BTZ} M. Banados, C. Teitelboim \& J. Zanelli,
          {\sl Phys. Rev. Lett.} {\bf 69}, 1849 (1992);
          M. Banados, M. Henneaux, C. Teitelboim \& J. Zanelli,
          {\sl Phys. Rev. D} {\bf 48}, 1506 (1993).
  \bibitem{Ross} R.B. Mann \& S.F. Ross,
  {\sl Phys. Rev. D} {\bf 47}, 3319 (1993).
  \bibitem{Cangemi}D. Cangemi, M. Leblanc \& R.B. Mann,
  {\sl Phys. Rev. D} {\bf 48}, 3606(1993)
  \bibitem{Jolien} J.D. Brown, J. Creighton and R.B. Mann,
  WATPHYS TH94/03, gr-qc 9405007.
  \bibitem{Price} R. H. Price, Phys. Rev. D {\bf 5}, 10, 2419 (1972).
\end{thebibliography}
\end{document}